\documentstyle[twoside]{article}

 at 8truept
\font\big=cmr10 at 12truept
\font\bigbf=cmbx10 at 14truept

\def\al{\alpha}
\def\be{\beta}
\def\ga{\gamma}
\def\de{\delta}
\def\ep{\epsilon}

\def\ka{\kappa}
\def\la{\lambda}

\def\ps{\psi}

\def\Ga{\Gamma}

\def\cl{{\cal L}}
\def\fr#1#2{{{#1} \over {#2}}}

\def\vev#1{\langle {#1}\rangle}

 \def\half{{\textstyle{1\over 2}}}
\def\frac#1#2{{\textstyle{{#1}\over {#2}}}} 
\def\lsim{\mathrel{\rlap{\lower4pt\hbox{\hskip1pt$\sim$}} 
\raise1pt\hbox{$<$}}}
\def\gsim{\mathrel{\rlap{\lower4pt\hbox{\hskip1pt$\sim$}} 
\raise1pt\hbox{$>$}}}
\def\sqr#1#2{{\vcenter{\vbox{\hrule height.#2pt 
\hbox{\vrule width.#2pt height#1pt \kern#1pt \vrule width.#2pt}
\hrule height.#2pt}}}}

\def\lrprt{\stackrel{\leftrightarrow}{\partial}}

\def\lrDmu{\stackrel{\leftrightarrow}{D_\mu}}

\newcommand{\beq}{\begin{equation}}
\newcommand{\eeq}{\end{equation}}
\newcommand{\bea}{\begin{eqnarray}}
\newcommand{\eea}{\end{eqnarray}}
\newcommand{\rf}[1]{(\ref{#1})}

\renewenvironment{thebibliography}[1]
{ \rm
\begin{list}{\arabic{enumi}.}
{\usecounter{enumi} \setlength{\parsep}{0pt} 
\setlength{\itemsep}{3pt} \settowidth{\labelwidth}{#1.} \sloppy
}}{\end{list}}

\begin{document}

\markboth{\bf D. COLLADAY}{\bf SPONTANEOUS VIOLATION OF LORENTZ SYMMETRY}
\pagestyle{myheadings}

\vglue 96 truept 

\baselineskip 16pt
\begin{flushright}
{\bigbf SPONTANEOUS VIOLATION OF LORENTZ AND CPT SYMMETRY \\}
\end{flushright}

\vglue 24pt

\renewcommand{\thefootnote}{\fnsymbol{footnote}}
\begin{flushleft}
{\big Don Colladay} \footnote{\noindent New College of the University of South
Florida,  Sarasota, FL, 34243.}
\end{flushleft}
\renewcommand{\thefootnote}{\arabic{footnote}}
\setcounter{footnote}{0}

\baselineskip=11pt

\vglue 24pt

\noindent
{\bf 1. INTRODUCTION}
\vglue 0.4 cm

The standard model as well as many of its modern day extensions preserves
Lorentz and CPT symmetry. 
In fact, symmetry under the Lorentz group is a basic assumption in 
virtually any fundamental theory used to describe elementary particle 
physics.  Under very mild assumptions,
the postulates of a point particle theory that preserves Lorentz 
invariance imply that CPT is also preserved \cite{cptthm}.

In this proceedings, I will discuss the construction of quantum field 
theories that break Lorentz and CPT symmetry.  There are both experimental and 
theoretical motivations to develop such theories.

Many sensitive experimental tests of Lorentz and CPT symmetry have 
been performed.
For example, high precision tests involving  
atomic systems
\cite{ptrap,bkr},
clock comparisons \cite{cc, klcc}, and neutral meson oscillations 
\cite{kexpt, colladayb}
provide stringent 
tests of Lorentz and CPT symmetry.
Recently, a pendulum with a net macroscopic spin angular momentum has been
constructed and used to investigate spin-dependent Lorentz and CPT violation
\cite{bkp}.
In the past, such experiments have placed  bounds on
phenomenological  parameters that lack any clear connection with the microscopic
physics  of the standard model. One motivation of constructing a theory in the
context of the standard  model that allows for Lorentz and CPT violation is the
desire to have a single theory within the context of  conventional quantum field
theory that could  relate various experiments and be used to motivate future
investigations.

This begs the question as to how such effects might arise naturally within the 
current framework of quantum field theory.
The main idea is that miniscule low-energy remnant effects that violate 
fundamental symmetries may arise in theories underlying the standard  model.  One
example is string theory in which nontrivial structure  of the vacuum solutions may
induce observable Lorentz and CPT violations \cite{kp1,kp2,kp3}.

Rather than attempting a construction based directly on a specific underlying 
model, such as string theory, we proceed using the generic mechanism of
spontaneous symmetry violation to implement the breaking.
Terms involving standard model fields that violate Lorentz and
CPT  symmetry are assumed to arise from a general spontaneous symmetry breaking
mechanism in which vacuum expectation values for tensor fields are generated in the
underlying theory
\cite{ck1}. The 
approach then is to construct all possible terms that can arise 
through spontaneous symmetry breaking that are consistent with the gauge 
invariance of the standard model and power-counting renormalizability.
These conditions are imposed on the model to limit the deviation from the
conventional standard model by preserving gauge symmetries and renormalizability.
It is somewhat analogous to imposing R-parity in supersymmetry to eliminate 
pesky lepton number violating interactions.

The resulting terms lead to modified field equations that can be 
analyzed within the context of conventional quantum field theory.  
In this proceedings, I will develop the modified Feynman rules for a model theory 
and will explore some possible consequences for 
quantum electrodynamics.  Other topics in Lorentz and CPT violation including
a detailed analysis of causality and stability issues \cite{kproc} and an 
investigation of effects on neutrino oscillations \cite{whis} are also being 
discussed at this meeting.

\vglue 0.6 cm
\noindent
{\bf 2. SPONTANEOUS BREAKING OF LORENTZ SYMMETRY}
\vglue 0.4 cm

In this section, a general spontaneous symmetry breaking mechanism is applied to the
fermion sector to generate an example of the types of interactions that arise.
The conventional mechanism of this type occurs in theories that 
contain scalar field potentials with nontrivial minima, such as 
the conventional Higgs mechanism of the standard model in which Yukawa 
couplings generate the
fermion masses after spontaneous symmetry breaking of the scalar Higgs field.
In conventional theories of this sort, internal symmetries of the original 
Lagrangian such as gauge invariance may be violated, but 
Lorentz symmetry is always maintained.

The key element in preserving Lorentz invariance when symmetry is broken 
spontaneously is the fact
that a Lorentz scalar obtains an expectation value.
Spontaneous Lorentz breaking may occur in a fundamental theory 
containing a potential for a tensor field that has nontrivial minima.
For example, consider a lagrangian 
describing a fermion $\psi$ and a tensor $T$ of the form
\beq
{\cal{L}} = {\cal{L}}_{0} - {\cal{L}}^{\prime} 
\quad ,
\eeq
where
\beq
{\cal L}^{\prime} \supset \fr {\la}{M^k} {T} \cdot \overline{\psi} 
\Ga (i \partial )^k \psi + \rm{h.c.} + V(T)
\quad .
\eeq
In this expression, $\la$ is a dimensionless coupling constant, 
$M$ is some heavy mass scale of the underlying theory,
$\Ga$ denotes a general gamma matrix element in the Dirac algebra,
and $V(T)$ is a potential for the tensor field (indices are
suppressed for notational simplicity).
The potential $V(T)$ is assumed to arise in a more fundamental theory underlying
the standard model.  
Note that terms contributing to $V(T)$ are precluded from
conventional renormalizable four-dimensional field theories,
making this type of violation impossible.
However, these terms are naturally generated in the low-energy
limit of more general theories such as string theory \cite{kp1,kp2}.

If the function $V(T)$ has nontrivial minima, 
a nonzero expectation value of $T$ will be generated in the vacuum.
The lagrangian will then contain a term of the form
\beq
{\cal L}^{\prime} \supset \fr {\la}{M^k} \vev{T} \cdot \overline{\psi} 
\Ga (i \partial )^k \psi + \rm{h.c.} 
\quad ,
\label{vevt}
\eeq
that is bilinear in the fermion fields and can violate Lorentz 
invariance and various discrete symmetries C, P, T, CP, and CPT.

\vglue 0.6 cm
\noindent
{\bf 3. RELATIVISTIC QUANTUM MECHANICS} 
\vglue 0.4 cm

To illustrate the techniques for treating terms of the form generated
in Eq.\rf{vevt},
we examine a subset of all the possible terms.
An example applicable to the standard model fermions is furnished by 
the choice $k=0$ (no derivatives) and $\Ga \sim \ga^{\mu}$ or 
$\Ga \sim \ga^{5}\ga^{\mu}$, the most general nonderivative 
terms that violate CPT
symmetry. With these restrictions, the model lagrangian for a single 
fermion $\psi$ becomes
\beq
{\cal L} = \fr i 2 \overline{\ps} \ga^{\mu} \lrprt_\mu \psi 
- a_{\mu} \overline{\psi} \ga^{\mu} \psi 
- b_{\mu} \overline{\psi} \ga_5 \ga^{\mu} \psi 
- m \overline{\psi} \psi
\quad ,
\label{modlag}
\eeq
where the parameters $a_{\mu}¥$ and $b_{\mu}¥$ are real constant 
coefficients that denote the tensor expectation values and coupling 
constants that are present in \rf{vevt}.

Several features of this theory can be immediately deduced from the
structure of the lagrangian.  First, the lagrangian is hermitian and 
therefore preserves
probability.  This  means that conventional quantum mechanics can be used to
evolve the particle states in time.
The model lagrangian is invariant under translations and U(1) 
gauge transformations which leads to conservation of energy, momentum, 
and charge.
The resulting Dirac equation
\beq
(i \ga^{\mu} \partial_{\mu} - a_\mu \ga^\mu - 
b_{\mu} \ga_5 \ga^\mu - m) \psi = 0
\quad .
\label{de}
\eeq
obtained by minimizing the variation of the action with respect to
the fermion field is linear in
$\psi$.  Equation \rf{de} can be solved exactly using the plane-wave 
solutions
\beq
\psi(x) = e^{\pm i p_{\mu} x^{\mu}} w(\vec{p}) 
\quad ,
\eeq
where $p^0(\vec{p}) \equiv E(\vec{p})$ is the energy (defined as the
magnitude of the eigenvalue of 
the hamiltonian acting on the state) determined by  setting the determinant of the
matrix acting on
$w(\vec{p})$ equal to  zero. 

The general form of the resulting dispersion relation is complicated, 
so we content ourselves here by investigating the special case $\vec{b} = 0$. The
exact solutions for the energies are found to be
\beq
E_{+}(\vec{p}) = \left[ m^2 + (|\vec{p} - \vec{a}| \pm b_0)^2 \right]^{1/2} 
+ a_0 \quad ,
\eeq
\beq
E_{-}(\vec{p}) = \left[ m^2 + (|\vec{p} + \vec{a}| \mp b_0)^2 \right]^{1/2} 
- a_0 \quad .
\eeq
Some interesting consequences of the breaking is apparent.
Note that the conventional energy degeneracy of the fermion 
and antifermion states is broken by $a_{\mu}$ 
while $b_{0}$ splits the degeneracy of the helicities.
These energy splittings are indicative of the effect of the general Lorentz
violating terms in the standard model extension. The corresponding spinor solutions
form an orthogonal basis of states as a result of the hermiticity of the hamiltonian.

An interesting feature of the above dispersion relations is the 
modified relationship that exists between the velocity of a wave packet 
and its corresponding momentum.
For instance, a wave packet formed from a superposition of positive 
helicity fermions with a four-momentum $p^{\mu}=(E,\vec{p})$
has a corresponding expectation value for the velocity operator 
$\vec{v} = i[H,\vec{x}] = \ga^0 \vec{\ga}$ of
\beq
\vev{\vec{v}} = 
\vev{ \fr {(|\vec{p} - \vec{a}| - b^0)} {(E - a^0)}
         \fr {(\vec{p} - \vec{a})} {|\vec{p} - \vec{a}|} }
         \quad .
\eeq
Note that the velocity of the packet and the conserved momentum are not
in the same direction.
Examination of the velocity using a general nonzero $b_{\mu}$
reveals that $|v_{j}| < 1$,
and that the limiting velocity as $\vec{p} \rightarrow \infty$ is 1.
This implies that effects due to the CPT violating terms are mild 
enough to preserve causality.\footnote{Note that while causality is
preserved,  there can be problems with  stability at energies nearing the heavy mass
scale of the underlying theory \cite{kl}.   Also see Kosteleck\'y, these
proceedings.} This will be verified independently from the perspective of 
quantum field theory that will now be discussed.

\vglue 0.6 cm
\noindent
{\bf 4. CONSTRUCTION OF THE FREE FIELD THEORY}
\vglue 0.4 cm

The approach to quantization taken here is similar to the conventional 
one in which the quantization conditions on the fields are deduced from 
the requirement of positivity of the energy.
The wave function $\psi$ is expanded in terms of its four solutions as
\begin{eqnarray}
\ps (x) & = & \int \fr {d^3 p} {(2 \pi )^3} \sum_{\al = 1}^{2} \left[
\fr m {E_u^{(\al)}} b_{(\al)} (\vec{p})
e^{-i p_u^{(\al)} \cdot x} u^{(\al)} (\vec{p}) \right. \nonumber \\ 
&& \left. \qquad \qquad \qquad \qquad
+ \fr m {E_v^{(\al)}} d^*_{(\al)} (\vec{p}) 
e^{i p_v^{(\al)} \cdot x} v^{(\al)} (\vec{p}) \right] 
\quad ,
\end{eqnarray}
and is promoted to an operator acting on a Hilbert space 
of basis states.

Translational invariance is used to define the conserved energy and momentum 
as
\beq
P_\mu = \int d^3 x \Theta^0_{\ \mu} = 
\int d^3 x \frac 1 2 i \overline{\psi} \ga^0 \lrprt_\mu \psi
\quad .
\label{emom}
\eeq
The time component $P_{0}$ is interpreted as the energy (after normal 
ordering of the operators) and is positive definite (for $|a^{0}| <m$)
provided the following anticommutation relations are imposed:
\begin{eqnarray}
\{b_{(\al)} (\vec{p}), b^{\dagger}_{(\al^{\prime})} (\vec{p}^{~\prime}) \} 
& = & (2 \pi)^3
\fr {E_u^{(\al)}} {m}
\de_{\al \al^{\prime}}
\de^3 (\vec{p} - \vec{p}^{~\prime})
\quad ,
\nonumber \\
\{d_{(\al)} (\vec{p}), d^{\dagger}_{(\al^{\prime})} (\vec{p}^{~\prime}) \} 
& = & (2 \pi)^3
\fr {E_v^{(\al)}} {m}
\de_{\al \al^{\prime}}
\de^3 (\vec{p} - \vec{p}^{~\prime})
\quad .
\end{eqnarray}
The resulting equal-time anticommutators of the fields are
\begin{eqnarray}
\{ \psi_{\al}(t,\vec{x}), \psi_{\be}^{\dagger}(t,\vec{x}^{\prime})\} 
& = & \de_{\al \be} \de^{3} (\vec{x} - \vec{x}^{\prime})
\quad , 
\nonumber \\ 
\{ \psi_{\al}(t,\vec{x}), \psi_{\be}(t,\vec{x}^{\prime})\} & = & 0 
\quad ,
\nonumber \\
\{ \psi_{\al}^{\dagger}(t,\vec{x}), \psi_{\be}^{\dagger}(t,\vec{x}^{\prime})\} 
& = & 0
\quad .
\end{eqnarray}
These relations show that conventional Fermi statistics remain 
unaltered by the CPT violation.

The conserved charge $Q$ and conserved momentum $P_{\mu}$ 
are now computed explicitly as
\begin{eqnarray}
Q & = & \int \fr {d^3 p} {(2 \pi)^3}
\sum_{\al = 1}^2 \left[
\fr m {E_u^{(\al)}}
b^{\dagger}_{(\al)} (\vec{p}) b_{(\al)} (\vec{p}) - \fr m {E_v^{(\al)}}
d^{\dagger}_{(\al)} (\vec{p}) d_{(\al)} (\vec{p}) \right] \quad ,\\
P_{\mu} & = & \int \fr {d^3 p} {(2 \pi)^3} \sum_{\al = 1}^2 \left[
\fr m {E_u^{(\al)}} p^{(\al)}_{u \mu}
b^{\dagger}_{(\al)} (\vec{p}) b_{(\al)} (\vec{p}) \right. 
\nonumber \\ 
& & \qquad \qquad \qquad \qquad \qquad \left.
+ \fr m {E_v^{(\al)}} p^{(\al)}_{v \mu}
d^{\dagger}_{(\al)} (\vec{p}) d_{(\al)} (\vec{p}) \right]
\quad .
\end{eqnarray}
From these expressions, it is observed that the charge of the fermion is 
unperturbed and the energy and momentum satisfy the same energy-momentum 
relations that were found using relativistic quantum mechanics.

To preserve causality, it is necessary that the anticommutation relations 
of the fermion fields at unequal times are zero for spacelike separations.  
Explicit integration for the special case of $\vec{b}=0$ reveals that
\beq
\{\psi_{\al}(x), \overline{\psi}_{\be}(x^{\prime})\} = 0
\quad ,
\eeq
for spacelike separations $(x - x^{\prime})^2 < 0$.
This result indicates that physical observables separated by spacelike 
intervals will in fact commute.
This agrees with our previous results regarding the velocity 
obtained using the relativistic quantum mechanics approach.
An analysis of causality and stability issues for other Lorentz- and
CPT-violating terms in the fermion sector of the standard model extension 
has recently been performed 
\cite{kl}.  Some similar issues pertaining to causality in the photon sector have
subsequently been addressed
\cite{klink}.

Next, the issue of extending this free field theory to interacting 
theory is addressed.
Much of the conventional formalism developed for perturbative 
calculations in conventional interacting field theory carries over 
essentially unchanged to the present case.
The asymptotic in and out states are defined as in the usual 
case using the free field solutions.  The LSZ reduction procedure is then used to
express the transition-matrix elements in terms of Green's functions for the theory.
Dyson's formalism is then used to express the time-ordered products of the 
interacting fields in terms of the asymptotic fields.
Wick's theorem remains unaffected by the modifications.

A central result is that the usual Feynman rules apply provided that 
the Feynman propagator is modified to
\beq
S_F(p) = \fr i {p_{\mu} \ga^{\mu} - a_{\mu} \ga^{\mu} - 
b_{\mu} \ga_5 \ga^{\mu} - m +i \epsilon}
\quad ,
\eeq
and the exact spinor solutions of the modified free fermion theory 
are used on external legs.
The main reason that conventional techniques apply seems to be due to the fact that 
the Lorentz  violating modifications are linear in the fermion fields.

\vglue 0.6 cm
\noindent
{\bf 5. QED EXTENSION AND THE PHOTON}
\vglue 0.4 cm

In this section, some implications of Lorentz breaking for 
photon propagation are investigated.
The conventional QED lagrangian is 
\beq
\cl^{\rm QED}_{\rm electron} = 
\half i \overline{\ps} \ga^\mu \lrDmu \ps 
- m_e \overline{\ps} \ps
- \frac 1 4 F_{\mu\nu}F^{\mu\nu}
\quad ,
\eeq
where $\psi$ is the electron field, $m_{e}$ is its mass,
and $F^{\mu\nu}$ is the photon field strength tensor.
When all possible Lorentz-violating contributions from spontaneous 
symmetry breaking consistent with gauge invariance and power-counting 
renormalizability are introduced into the standard model,
the resulting modifications to the photon sector are \cite{ck1}
\beq
\cl^{\rm CPT-even}_{\rm photon} =
-\frac 1 4 (k_F)_{\ka\la\mu\nu} F^{\ka\la}F^{\mu\nu}
\quad ,
\eeq
and
\beq
\cl^{\rm CPT-odd}_{\rm photon} =
+ \half (k_{AF})^\ka \ep_{\ka\la\mu\nu} A^\la F^{\mu\nu}
\quad .
\eeq
The parameters $k_{F}$ and $k_{AF}$ are fixed background fields related to 
vacuum expectation values of tensors coupled to the photon in the underlying theory.
These two couplings are even and odd under CPT respectively.
The CPT-odd terms have been treated in detail elsewhere \cite{ck1, cfj}.
Here the special case of
$(k_{AF})^\mu = 0$ (no CPT-odd piece), and 
$(k_F)_{0j0k} = - \half \be_j\be_k$ is examined for the sake of a specific example.

The resulting modifications to the Maxwell equations are linear just 
as the modified Dirac equation is in the fermion case.
Plane waves can therefore be used to solve the modified equations of motion.
A solution exists provided $p_{\mu}$ satisfies 
\begin{eqnarray}
(p_o)^2 & = & 0 
\quad , \\
(p_e)^2 & = & - \fr { (\vec{\be}\times \vec p_e)^2}
{1 + \vec{\be}^2}
\quad ,
\end{eqnarray}
where $p_{o}$ denotes an ordinary mode 
and $p_{e}$ denotes an extraordinary mode of photon propagation.
The ordinary mode propagates as a conventional photon, while
the extraordinary mode has a modified dispersion relation.

For the special direction of propagation for which $\vec{\be} \cdot \vec{p} = 0$, 
the ordinary mode is polarized with $\vec{A}_{o}$ along the direction 
of $\vec{p} \times \vec{\be}$ while the extraordinary mode 
$\vec{A}_{e}$ is polarized along $\vec{\be}$.
Both polarizations are perpendicular to the momentum vector of the wave
$\vec{p}$.
The group velocities of wave packets $\vec v_g \equiv \vec \nabla_p p^0$ 
are calculated as
\beq
\vec v_{g,o} = \hat p 
\quad , \quad
\vec v_{g,e} = 
\fr 1 {\sqrt{1 + \vec{\be}^2}} ~ \hat p 
\quad .
\eeq
The extraordinary mode is seen to travel with a modified velocity that is 
slightly less than the velocity of the ordinary mode.
As a result, an initially plane polarized wave will in general become 
elliptically polarized after traveling a distance
\beq
r \simeq \fr {\pi} 
{2 \left( \sqrt{1 + \vec{\be}^2} - 1 \right) p^0}
\simeq \fr {\pi} {\vec{\be}^2 p^{0}}
\quad ,
\eeq
where the approximation holds for $\vec{\be}^2 \propto k_{F} << 1$.
The magnetic field behaves analogously.
Terms of this form can also have implications for photon birefringence.
In particular they can contribute to polarization rotation from
distant quasars \cite{pk}.

\vfill\eject

\vglue 0.6 cm
\noindent
{\bf REFERENCES}
\vglue 0.4 cm

\end{document}